\documentclass[twocolumn]{aa}
\usepackage{epsf,txfonts}
\begin{document}
\authorrunning{Drury et al.}
\title{Particle acceleration in supernova remnants, the
Bell-Lucek hypothesis and the cosmic ray ``knee''.}
\author{L. O' C. Drury
\inst{1}, 
E. van der Swaluw
\inst{1,2} 
and O. Carroll
\inst{1,3}}
\institute{Dublin Institute for Advanced Studies, 5 Merrion Square, Dublin 2, 
Ireland
\and
FOM-Institute for Plasma Physics Rijnhuizen, P.O. Box 1207, 3430 BE Nieuwegein, The Netherlands 
\and
Physics Department, University College Cork, Cork, Ireland
}
\offprints{L. O'C. Drury, {\tt ld@cp.dias.ie}}
\date{}

\abstract{Young supernova remnants are thought to be the sites where cosmic
ray acceleration occurs by the mechanism of diffusive shock acceleration. The 
maximum energy gained in this process is conventionally extimated to have a value close to,
but distinctly below, the ``knee'' ($\sim 10^{15}$ eV) of the cosmic-ray spectrum. Bell \& Lucek (2001) 
have suggested that the generated cosmic rays simultaneously amplify the magnetic field around the
supernova remnant shock to many times its pre-shock value. In this case the 
acceleration rate may be significantly increased and protons can easily reach 
energies up to $10^{17}$ eV. We use a ``simplified'' box model
incorporating the magnetic field amplification suggested by Bell \& Lucek to investigate the 
resulting modifications of the cosmic-ray spectrum. The model predicts a 
spectral break at high energies, close to the ``knee'' region, and in good accordance with 
observations.
}

\maketitle
\section {Introduction}

Bell \& Lucek (2001) and Lucek \& Bell (2000) have presented numerical
simulations suggesting that the conventional process of particle
acceleration in shocks, generally called diffusive shock acceleration,
may also result in substantial amplification of the highly tangled
magnetic field around the shock. The unusual strong magnetic field
inferred in Cas A (e.g. Vink \& Laming 2003) seems to confirm this
prediction. We note in passing that this hypothesis provides a
concrete physical mechanism for one of the many ideas of the late F
Hoyle who speculated (Hoyle 1960) that strong interstellar shocks
might convert mechanical energy into roughly equal amounts of magnetic
field energy, cosmic ray energy and thermal energy. It is also closely
related to the ``plastic deformation'' of the magnetic field discussed
qualitatively by V\"olk and McKenzie (1981).  Bell \& Lucek (2001)
suggest that this may enable strong supernova driven shocks in the
interstellar medium to accelerate protons to energies well beyond what
are conventionally held to be the maximum realistically attainable
energies of a few times $10^{14}\,\rm eV$ (e.g. Lagage and Cesarsky,
1983).

The Bell-Lucek hypothesis is of great interest because it is one of
the few suggestions as to how the cosmic ray particles of energies at
and above the ``knee'', located at a few times $10^{15}\,\rm eV$,
could be made in the Galaxy. It is therefore important to determine
what the expected form of the source energy spectrum would be and
whether the slight break in the observed spectrum at the ``knee'' can
be accounted for in this way.  There are also interesting implications
for gamma ray observations of SNRs with the next generation of imaging
atmospheric Cherenkov telescopes such as HESS.  For an initial
examination of this problem the so-called ``box'' models of particle
acceleration (Drury et al. 1999) offer enough accuracy and have the great 
advantage of computational simplicity.

\section{The Bell-Lucek hypothesis}

Bell \& Lucek point out that in the conventional diffusive shock
acceleration picture the energy density in the accelerated particles
at and near the shock front is extremely large, of order the total ram
pressure of the incoming plasma, and thus much larger than the energy
density of the magnetic field (the ratio of particle energy to field
energy is of order the square of the shock Alfven Mach number). The
standard treatment of resonant wave excitation, in which the
perturbations of the field are treated as Alfven waves to lowest order
and the interaction with the particles as a small perturbation, is
thus very questionable. With the support of numerical simulations and
simplified analytic models they suggest that in reality the field can
be highly distorted by the particle pressure and wound up to the point
where approximate equipartition holds.

If this amplified and distorted field is then used to estimate the
particle diffusion in the shock neighbourhood, acceleration to
substantially higher energies than in the conventional picture is
obviously possible.  Detailed estimates and simple dimensional
analysis agree that the maximum particle rigidity is given, to order
of magnitude, by the product of the field strength $B$, the shock
radius $R$ and the shock velocity $U$.  Thus, other things being
equal, the increase is directly proportional to the increase in the
field strength, which by the above argument is of order the Alfven
Mach number of the shock ${\cal M}_{\rm sh}$. This can easily be
${\cal M}_{\rm sh}\simeq 10^3$ for a young supernova remnant so that
the effect is potentially very significant; acceleration to rigidities
of a few $10^{17}\,\rm V$, rather than the $10^{14}\,\rm V$ normally
estimated, is easily possible.

Ptuskin and Zirakashvili (2003) have performed a sophisticated analysis of the
combined effects of field amplification and wave damping on particle
acceleration in supernova remnants and conclude that indeed the upper cut-off
to the accelerated spectrum can be quite strongly time-dependent. Our aim in this
short paper is to carry out a first exploratory analysis of the effect this has
on the overall spectrum and we therefore use the simple approach of assuming
an equipartition field although this is clearly a rather crude approximation.

\section{The ``box'' model}
To get a first estimate of the effect of such a dynamically generated
field on the acceleration we turn to the simplest treatment of shock
acceleration, the so-called ``box'' model. In this the accelerated
particles are assumed to be more or less uniformly distributed
throughout a region extending one diffusion length each side of the
shock, and to be accelerated upwards in momentum space at the shock
itself with an acceleration flux
\begin{equation}
\Phi(p) = {4 \pi\over3} p^3 f(p) \left(U_1 - U_2\right),
\end{equation} 
per unit surface area
where $U_{1}$ and $U_{2}$ are the upstream and downstream velocity and $f(p)$ 
is the phase space density of the accelerated particles (assumed to have an \
almost isotropic distribution).
If the diffusion
length upstream is $L_1$, and that downstream is $L_2$, then
\begin{equation}
L_1 \approx {\kappa_1(p)\over U_1},\qquad L_2 \approx {\kappa_2(p)\over U_2},
\end{equation}
where $\kappa_1$ and $\kappa_2$ are the upstream and downstream
diffusion coefficients. To a first approximation we assume that both
$L_1$ and $L_2$ are small relative to the radius of the shock and that
we can neglect effects of spherical geometry (in fact it is not too
difficult to develop a spherical box model, but it unnecessarily
complicates the argument) so that the box volume is simply
$A(L_1+L_2)$ where $A$ is the surface area of the shock.
The basic ``box'' model equation is then simply a conservation
equation for the particles in the box; the rate at which the number in
the box changes is given by the divergence of the acceleration flux in
momentum space plus gains from injection and advection and minus
advective losses to the downstream region.
\begin{eqnarray}
{\partial\over\partial t} \left[ A (L_1 +  L_2) 4\pi p^2 f(p) \right]
&+& A{ \partial \Phi\over\partial p} = A Q(p) \nonumber\\
&+& A F_1(p) - A F_2(p),
\end{eqnarray}
where $Q(p)$ is a source function representing injection at the shock
(only important at very low energies), $F_1$ is a flux function
representing advection of pre-existing particles into the system from
upstream (normally neglected) and $F_2$ is the flux of particles
advected out of the system and carried away downstream.  The only
complication we have to consider is that the box is time-dependent,
with flow speeds, shock area and diffusion lengths all changing.

The escaping flux is determined simply by the advection across the 
downstream edge of the box, that is
\begin{equation}
F_2(p) = 4 \pi p^2 f(p) \left(U_2 - {\partial L_2\over\partial
  t}\right),
\end{equation}
where we have to explicitly allow for the time varying size of the
downstream region.
Substituting this expression for $F_2(p)$ and neglecting the advection of prexisting particles 
(the $F_1(p)$ term) the box equation
simplifies to:
\begin{eqnarray}
{1\over A} {\partial A\over\partial t} \left( L_1+L_2 \right) f
&+& {\partial L_1\over\partial t} f
+\left(L_1+L_2\right){\partial f\over\partial t} 
+ U_1 f \nonumber\\
&+&\left(U_1 - U_2\right) {p\over 3} {\partial f\over\partial p} =
{Q\over 4\pi p^2}.
\end{eqnarray}

\section{Incorporating the Bell-Lucek effect in the ``box'' models}

Partial differential quations of this form always reduce, by the
method of characteristics, to the integration of two ordinary
equations, one for the characteristic curve in the $(p,t)$ plane
\begin{equation}
\label{char}
{{\rm d}\,p\over {\rm d}\,t} = {U_1 - U_2 \over L_1 + L_2} {p\over3},
\end{equation}
and one for the variation of $f$ along this curve
\begin{eqnarray}
\left(L_1 + L_2\right) {{\rm d}\,f\over {\rm d}\,t} 
&+& f\left[\left(L_1+L_2\right){1\over A}{\partial A\over\partial t}
+{\partial L_1\over\partial t} + U_1\right] \nonumber\\
&=& {Q\over 4\pi p^2}.
\end{eqnarray}
Apart from at the injection momentum                                                                
$Q=0$ and we can write the
above equation as
\begin{equation}
\label{PDE}
{{\rm d}\, \ln f\over {\rm d}\, t}
= - {1\over A}{\partial A\over\partial t}
  - {1\over L_1+L_2} {\partial L_1\over \partial t}
  - {U_1\over L_1 + L_2}.
\end{equation}
But the
shock area $A$ is a function only of time so that 
\begin{equation}
{\partial A\over\partial t} = {{\rm d}\, A\over {\rm d}\, t},
\end{equation}
and, although the upstream diffusion length does depend on both time
and momentum, if we assume Bohm scaling for the two lengths so that
\begin{equation}
L \propto {\kappa\over U}  \propto {p v\over U B},
\end{equation}
(where $v$ is the particle velocity) we can write
\begin{equation}
{1\over L_1+L_2} {\partial L_1\over \partial t}
= - \vartheta {{\rm d}\,\ln(U_1 B_1)\over {\rm d}\, t},
\end{equation}
where
\begin{equation}
\vartheta = {L_1\over L_1 + L_2},
\end{equation}
(obviously $0<\vartheta<1$).
Finally, noting that
\begin{equation}
{U_1\over L_1 + L_2} = {3 U_1\over U_1 - U_2} {{\rm d}\,\ln p\over 
{\rm d}\, t},
\end{equation}
and assuming that the shock remains strong, which yields a fixed compression
ratio, we can simplify equation (\ref{PDE}) to 
\begin{equation}
{{\rm d}\, \ln f\over {\rm d}\, t} = - {{\rm d}\, \ln A\over {\rm d}\, t}
+ \vartheta {{\rm d}\,\ln(U_1 B_1)\over {\rm d}\, t}
-  {3 U_1\over U_1 - U_2} {{\rm d}\,\ln p\over {\rm d}\, t},
\end{equation}
which integrates trivially to relate the value of $f$ at the end of
one of the characteristic curves, say at the point $(p_1, t_1)$, to the
value at the start, say at $(t_0, p_0)$, as follows;
\begin{equation}
\label{int1}
{f(t_1, p_1)\over f(t_0, p_0)} =
\left(A(t_1)\over A(t_0)\right)^{-1} 
\left(U_1(t_1) B_1(t_1)\over U_1(t_0) B_1(t_0)\right)^\vartheta
\left(p_1\over p_0\right)^{-s}
\end{equation}
where
\begin{equation}
s = {3U_1\over U_1-U_2}
\end{equation}
is the standard exponent of the steady-state power-law spectrum associated with
shock acceleration.

This rather beautiful result shows how the standard test particle
power-law is modified by a combination of effects as the box volume
changes. As one would expect the amplitude varies inversely as the
shock area and also decreases if the upstream diffusion length (at
fixed energy) increases, but with an exponent between zero and one
determined by the ratio of the upstream diffusion length to the total
width of the diffusion region. It is very interesting that the result
is not simply a variation inversely as the box volume, which one would
naively expect from geometrical dilution. This reflects the
fundamental asymmetry between the upstream and downstream regions,
that upstream is empty outside the diffusion region whereas the entire
downstream region is filled with accelerated particles.

If we assume pure Bohm scaling the other diffential equation is also
integrable so that the problem is reduced entirely to quadratures (of
course only within the various approximations we are making; but still
a remarkable result).  Bohm scaling implies that the mean free path is
of order and proportional to the particle gyroradius, so that if the
particle charge is $e$
\begin{equation}
L_1 + L_2 \approx \alpha {p v \over 3 e B_1 U_1},
\end{equation} 
where $\alpha$ is a dimensionless parameter (probably of order ten).
Substituting in the equation of the characteristic (equation (\ref{char}))we get
\begin{equation}
v {{\rm d}\, p\over {\rm d}\, t} = {1\over\alpha} \left(U_1-U_2\right) U_1 e B_1,
\end{equation}
and noting the relativistic identity between kinetic energy $T$,
momentum $p$ and velocity $v$,
\begin{equation}
v = {{\rm d}\, T\over {\rm d}\, p},
\end{equation}
we can integrate this as
\begin{equation}
\label{kinen}
T_1 - T_0 = {e\over\alpha} \int_{t_0}^{t_1} \left(U_1-U_2\right)U_1 B_1 \,dt.
\end{equation}
For relativistic particles the kinetic energy and the momentum are
essentially interchangeable with $T= c \sqrt{p^2 + m^2 c^2} - m c^2
\approx c p$.

These two integrals (equations (\ref{int1})\& (\ref{kinen})) together
reduce the problem of calculating the final spectrum to that of
determining the initial amplitude $f(t_0, p_0)$ which in turn depends
on the injection rate and its time dependence.

\section{The injection rate}

There are two main approaches to the injection rate.  The simplest,
which is perhaps more consistent with the test particle approach, is
to simply parametrise it by assuming that some fraction of the
incoming thermal particles are ``injected'' as non-thermal particles
at some suitably chosen ``injection momentum'' which 
separates the thermal particle population from the non-thermal.  
In other words one writes
\begin{equation}
Q(p,t) = \eta(t) n_1 U_1 \delta(p-p_{\rm inj}(t)),
\end{equation} 
where $n_1$ is the upstream thermal particle number density, $\eta\ll 1$ is
the injection fraction, $p_{\rm inj}$ is the injection momentum and as
usual $\delta$ is Dirac's delta distribution.  It
should be clear that this is a parametrisation rather than a true
injection model, however it, or equivalent parametrisations, have been
very widely used, typically with $\eta$ taken to be a constant of
order $10^{-5}$ to $10^{-4}$ for protons and $p_{\rm inj}\approx 10
m_p U_1$ where $m_p$ is the proton mass.  However there is no real
justification for this apart from the fact that it seems to yield
reasonable results in many cases.

With the above parametrisation the distribution function just above
the injection energy can be simply determined by equating the
acceleration flux to the injection flux,
\begin{equation}
{4 \pi p_{\rm inj}^3\over 3} \left(U_1-U_2\right) f(p_{\rm inj}) 
= \eta n_1 U_1,
\end{equation}
giving
\begin{equation}
f(p_{\rm inj}) = {3\over 4 \pi p_{\rm inj}^3} {U_1\over U_1 - U_2} n_1
\eta.
\end{equation}

The second approach adopts the idea, which can be traced back to the
early work of Eichler, that the injection process is inherently
extremely efficient but that various feedback processes
operate to reduce it to the point where the accelerated particles 
carry a significant part of the energy dissipated in the
shock. Probably the most sophisticated modern version of this idea is
to be found in the papers by Malkov (eg Malkov 1998, Malkov et al 2000; see also
Kang et al 2002). This, or something
similar, is in fact required for the Bell-Lucek hypothesis to operate
because it requires the accelerated particle energy density to be
substantial and of order the ram pressure of the upstream flow. For
a standard spectrum close to $p^{-4}$ the energy is almost uniformly
distributed per logarithmic interval over the relativistic part of the
spectrum.  This suggests taking a reference momentum in the
mildly relativistic region, $p_0\approx m c$, and determining $f$ by
a relation of the form
\begin{equation}
{4\pi p_0^3\over 3} f(p_0) m c^2 \approx \beta \rho U_1\left(U_1-U_2\right)
\end{equation}
where $\beta$ is a number which depends logarithmically on the upper
cut-off and which for supernova remnants is probably somewhere between
$10^{-1}$ and $10^{-2}$. 

It is important to note that both injection models are
models for proton injection, the protons being the dynamically
dominant species. Unfortunately very little is known about the factors
controlling the injection of electrons and other minor species despite
their importance for diagnostic tests. It is also very probable that
the injection is nonuniform over the shock surface with a strong
dependence on the angle between the mean background field and the
shock normal (V\"olk et al, 2003).

\section{Application to the Sedov solution}

Let us now apply these ideas to the Sedov solution (also studied by
Taylor and von Neumann) for a strong spherical explosion in a cold gas
where the shock radius expands as $R\propto t^{2/5}$ and the shock velocity
decreases as $U\propto t^{-3/5}$.  On the Bell-Lucek hypothesis the
magnetic field also scales as the shock velocity, $B\propto t^{-3/5}$
and thus the characteristic acceleration curves (equation(\ref{kinen})) are 
given by
\begin{figure}
\epsfxsize=\hsize
\epsfbox{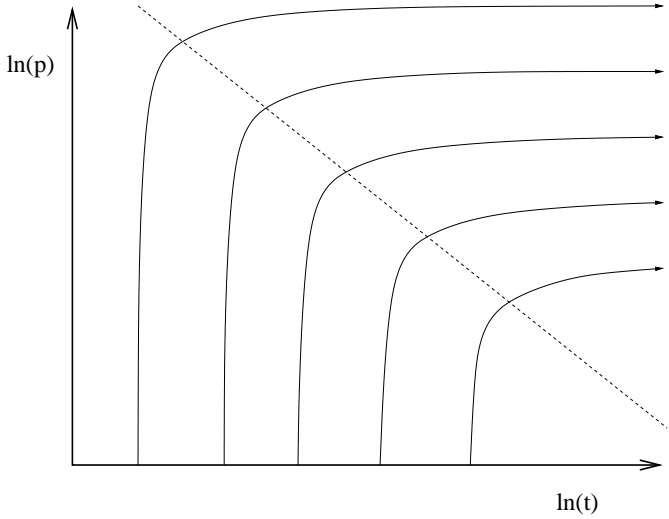}
\caption{The characteristic acceleration curves in the $t,p$ plane for a Sedov
blast wave and equipartition field amplification.  The dotted line indicates the
$p_1\propto t_0^{-4/5}$ asymptotic power-law relation between starting time and final momentum.}
\end{figure}
\begin{eqnarray}
T_1 - T_0 
&\propto& \int_{t_0}^{t_1} (U_1-U_2)U_1B dt\nonumber\\
&\propto& \int_{t_0}^{t_1} t^{-9/5} dt\nonumber\\
&\propto& t_0^{-4/5} - t_1^{-4/5}.
\end{eqnarray}
These curves, illustrated schematically in Fig.~1, all rise extremely
steeply, representing an initial phase of rapid acceleration, turn
over and then become asymptotically flat.  Physically it is clear
that, as the shock slows and the field drops, the high energy
particles cease to be significantly accelerated and simply diffuse
further and further in front of the shock. In fact in reality they
should probably be thought of as decoupling from the shock and forming
part of the general interstellar population at this point, but within
the box model they simply fill a steadily growing upstream region. We
will return to this point later.

A very important aspect of the curves is that they
uniquely relate final energies (or equivalently momenta) to starting
times. Asymptotically the relation is a simple power-law; for $T_1 \gg
T_0$ and $t_0 \ll t_1$ we have simply
\begin{equation}
p_1 \propto T_1 \propto t_0^{-4/5}, \qquad t_0 \propto p_1^{-5/4}.
\end{equation}
Using this we can now translate the dilution factors from
equation (\ref{int1}) to additional power-law terms in the final momentum. 
Explicitly, a given final momentum maps to a starting radius using a 
Sedov expansion-law:
\begin{equation}
R(t_0) \propto t_0^{2/5} \propto p_1^{-1/2},
\end{equation}
and thus the first term on the RHS of equation (\ref{int1}) translates to a 
$p_1^{-1}$ factor:
\begin{equation}
\label{RHS1}
\left(A(t_1)\over A(t_0)\right)^{-1}  \propto p_1^{-1}.
\end{equation}
The final momentum can also be mapped to a starting velocity, using again
a Sedov expansion-law:
\begin{equation}
U(t_0) \propto t_0^{-3/5} \propto p_1^{3/4}.
\end{equation}
The magnetic field, which on the Bell-Lucek hypothesis scales as
velocity, gives an additional $p_1^{3/4}$ factor, thus the second term on the
RHS of equation (\ref{int1}) scales as:
\begin{equation}
\label{RHS2}
\left(U_1(t_1) B_1(t_1)\over U_1(t_0) B_1(t_0)\right)^\vartheta\propto
p_1^{-3\vartheta/2}
\end{equation}
Furthermore we need to determine the initial amplitude of $f$ from an
injection model. First we use the $\eta$ parametrisation as discussed in 
section 5: 
\begin{equation} 
\label{inj1}
p_{\rm inj} = p_0 \propto U(t_0) \propto p_1^{3/4},
\end{equation}
together with
\begin{equation}
\label{inj2}
f_0 \propto \eta n p_0^{-3}.
\end{equation}
Finally assuming a strong shock which, using the Rankine-Hugoniot
conditions for a non-relativistic fluid, yields $U_1/U_2 = 4$ and 
$3U_1/(U_1-U_2) = 4$ and substituting the equations (\ref{RHS1}), 
(\ref{RHS2}), (\ref{inj1}) and (\ref{inj2}) into equation (\ref{int1}) 
one obtains a scaling-law for the particle distribution $f(p_1)$ at a 
fixed time $t_1$:
\begin{eqnarray}
f(p_1) &\propto& \eta n_0 p_0^{-3} A(t_0)
\left[U_1(t_0)B_1(t_0)\right]^{-\vartheta} \left(p_1\over
p_0\right)^{-4}\nonumber\\
&\propto& \eta  p_1^{3/4} p_1^{-1} p_1^{-3\vartheta/2} p_1^{-4}.
\end{eqnarray}
If $\eta$ is constant, the slope is steepened from the canonical value of 
$4$ to 
\begin{equation}
4.25+{3\vartheta\over 2}.
\end{equation}

\begin{figure}
\epsfxsize=\hsize
\epsfbox{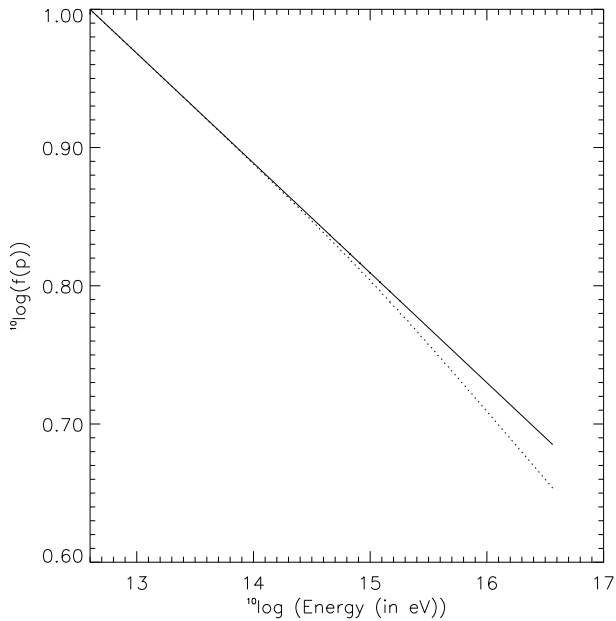}
\caption{Example of $f(t_1,p_1)$ (dotted line) from equation (15), where a 
Sedov-like shock has been to describe the velocity and $\vartheta=0.9$. 
The solid line is the standard $p^{-4}$ curve, the dotted line converges
towards the spectral index given by equation (34).}
\end{figure}

If we use the alternative ``equipartition'' argument as an injection model,
meaning that $f_0$ should be dynamically determined in the mildly relativistic 
region, we have $p_0\approx m c$ independent of $p_1$ and 
\begin{equation}
f(p_0) \propto U_1^2
\propto r^{-3} \propto p_1^{3/2},
\end{equation}
which gives the following scaling-law for the particle distribution:
\begin{equation}
f(p_1) \propto p_1^{3/2} p_1^{-1} p_1^{-3\vartheta/2} p_1^{-4}.
\end{equation}
In this approach the slope is given by
\begin{equation}
4 + {3\vartheta - 1\over2}.
\end{equation}
It is interesting that because of the strong injection at early times
this model can even, if $\vartheta<1/3$, lead to a slight flattening
of the spectrum. However, especially at high energies, it is unlikely
that the upstream diffusion region could be so small and a modest
steepening of the spectrum is more likely.

These results refer of course only to the asymptotic behaviour of the high
energy part of the spectrum. As $p_1$ is decreased there comes a point
where $t_0$ is no longer small relative to $t_1$. At this point all
values of the final momentum map down to a small approximately
constant region and the spectrum becomes simply the standard $p^{-4}$
spectrum.  This break occurs at the point to which efficient acceleration is possible at
that stage in the remnant evolution, and decreases as the the remnant ages. 
Shock acceleration will terminate when the shock is beginning to weaken and
the amplified field is only a few times the ambient field, which for ambient
fields of a few $\mu$Gauss and typical SNR parameters places the break
exactly in the ``knee'' region. 

Figure 2 shows an example of the total spectrum for a Sedov-like shock,
which was obtained by using equation (15).The injection mechanism for this 
example was taken according to the equations (22) and (23) and 
$\vartheta =0.9$. The spectrum clearly shows the smooth transition from the 
standard $p^{-4}$ spectrum to the asymptotic spectral index given by
equation (34).

\section{Conlcusions}

We have applied the Bell-Lucek hypothesis to a Sedov-like
shock, using a simplified Box model. We showed that such a 
model exhibits a spectral break at an energy determined by 
the current acceleration cut-off
below which one observes the standard shock acceleration spectrum, but
above which a slightly different power-law continues to higher
energies. For older remnants the break is expected to be in the
``knee'' region at rigidities of order $10^{15}\,\rm V$. The key point
is that the dynamical field amplification both increases the maximum
attainable energy and makes it a relatively strongly time-dependent
quantity unlike the situation with no field amplification where, as is
well known, there is only a very weak dependence of the cut-off energy
on the remnant age, at least during the Sedov phase (the much more
sophisticated analysis by Ptuskin and Zirakashvili, 2003, is relevant
here).

While the spectrum at the shock is the more relevant quantity from the
point of view of gamma-ray tests, in the context of cosmic ray
propagation theory what one would really like is the effective source
spectrum of an individual supernova integrated over its history.
This, while obviously related to the spectrum discussed here, is a
somewhat different quantity and harder to evaluate.  However it is
clear that here also one should expect a broken power-law with a
relatively small break in the exponent at a rigidity corresponding to
acceleration in a mildly amplified Galactic field.

Obviously further and more detailed work is needed, but it is very
encouraging that even such a simple model can produce spectra
remarkably close to the inferred cosmic ray source spectrum through
the ``knee'' region. In fact we are not aware of any other
acceleration model that can naturally produce a break of the right
magnitude (about 0.5 in the exponent) at the right position (modulo
major uncertainties in interstellar propagation at these energies of
course).

\end{document}